\documentclass[conference]{IEEEtran}
\PassOptionsToPackage{bookmarks=false}{hyperref}
\ifCLASSOPTIONcompsoc
 \usepackage[nocompress]{cite}
\else
 \usepackage{cite}
\fi
\usepackage{blindtext, graphicx}
\usepackage{pbox}
\usepackage{forest}
\usepackage{listings}
\usepackage{algorithm}
\usepackage[noend]{algpseudocode}
\usepackage{caption}
\usepackage{pdfcomment}
\usepackage{tcolorbox}
\usepackage{balance}
\usepackage{amsmath}
\usepackage[numbers]{natbib}
\usepackage{cite}
\usepackage{flexisym}
\algdef{SE}[SUBALG]{Indent}{EndIndent}{}{\algorithmicend\ }%
\algtext*{Indent}
\algtext*{EndIndent}
\renewcommand\IEEEkeywordsname{Keywords}

\newboolean{showcomments}
\setboolean{showcomments}{true} 
\ifthenelse{\boolean{showcomments}}
{ \newcommand{\mynote}[3]{
    \fbox{\bfseries\sffamily\scriptsize#1}
    {\footnotesize$\blacktriangleright$\textsf{\color{#3}{#2}}$\blacktriangleleft$}}}
{ \newcommand{\mynote}[3]{}}

\usepackage{array}

\usepackage[T1]{fontenc}

\begin{document}

 
 \title{Benchmarking and Performance Modelling of MapReduce Communication Pattern
 }

\author{\IEEEauthorblockN{Sheriffo Ceesay, Adam Barker and Yuhui Lin}
\IEEEauthorblockA{
University of St Andrews, UK\\
Email: \{sc306, adam.barker, yl205 \}@st-andrews.ac.uk
}

}

\maketitle

%
\begin{abstract}

Understanding and predicting the performance of big data applications running in the cloud or on-premises could help minimise the overall cost of operations and provide opportunities in efforts to identify performance bottlenecks. The complexity of the low-level internals of big data frameworks and the ubiquity of application and workload configuration parameters makes it challenging and expensive to come up with comprehensive performance modelling solutions. 
In this paper, instead of focusing on a wide range of configurable parameters, we studied the low-level internals of the MapReduce communication pattern and used a minimal set of performance drivers to develop a set of phase level parametric models for approximating the execution time of a given application on a given cluster. Models can be used to infer the performance of unseen applications and approximate their performance when an arbitrary dataset is used as input. Our approach is validated by running empirical experiments in two setups. On average the error rate in both setups is $\pm$10\% from the measured values.
\end{abstract}

%
%


%

%

%
\maketitle
\section{Introduction}
\begin{figure}[h!]
 \vspace{-2mm}
 \includegraphics[width=0.35\textwidth]{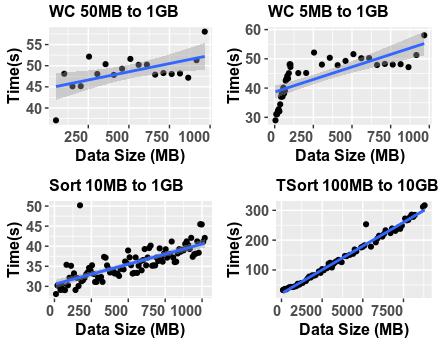}
 \centering
 \caption{\textbf{Execution time for varying datasets for MapReduce WordCount, TeraSort and Simple-Sort Programs }}
 \label{joblevel}
\end{figure}

Traditional data storage and processing like Relational Database Management Systems (RDBMS) by design are inefficient and rigid to store and handle big data. Over the years researchers have developed big data processing frameworks and storage systems to handle these challenges. 
Like any new system, it is crucial to model and understand its performs under various conditions. This could be a tedious and challenging task because big data application under the hood runs on frameworks comprising of complex and complicated pipeline of phases and operations. For example, applications deployed using MapReduce \cite{dean2008mapreduce} and Apache Spark \cite{zaharia2012resilient} frameworks go through several computation or communication stages. Phases and stages on a basic level are the transitions that the data goes through while being processed. More details on MapReduce phases are discussed in Section \ref{background}. Considering this challenge, we argue that a phase by phase modelling approach can gain a better understanding of the performance drivers of such frameworks. Some of the previous works in this area focuses only on high-level performance drivers like data size \cite{glushkova2017mapreduce,herodotou2011hadoop}. This approach may work well for some applications. For example, from sort operations shown in Figure \ref{joblevel}, we can see a clear linear relationship between the data size and the time taken to complete the execution. In this case, the input data size is equal to the output data size. However, as evident word count (WC) plots, the relationship between processing time and input data size is non-trivial and therefore there need for more investigation. The amount of data that passes through the various phases of the MapReduce pattern is crucial to its overall performance. This leads us to dig further into the low-level details of the framework. 
This work can also serve as a foundation for performance modelling of big data frameworks like Apache Spark, a general purpose computation engine that does of superset of what MapReduce does. The findings could be useful in size-based schedulers to run small jobs even when the cluster is loaded with long running and expensive jobs.

The research questions we seek to address in this paper are: 
\begin{itemize}
    \item RQ0: Considering the ubiquity of MapReduce workloads, can we use a generic approach to model the performance of MapReduce applications?
    \item RQ1: Considering how expensive it is to improvise a production size Hadoop cluster, can we identify major performance drivers for the MapReduce framework and use them to model the performance of applications?
    \item RQ2: Can we measure how good the effect of RQ1 would have on an arbitrary cluster?
\end{itemize}
In this paper, we develop a parametric model to estimate execution time of a given MapReduce application for a given cluster of nodes running YARN ~\cite{vavilapalli2013apache} (MapReduce 2.0 or greater). To achieve this, we first study and identify the performance drivers of each generic phase of the MapReduce application execution life cycle. This is achieved by executing generic MapReduce workloads to collect relevant performance metrics from logs. The combination of varying modelling approaches and cross-validation methods are used to obtain the models with their respective parameters. 
We have also developed a proof of concept application to test the models with different datasets. This links for the application and source code are is available at:

\noindent \href{https://sc306.host.cs.st-andrews.ac.uk/cbm/}{https://sc306.host.cs.st-andrews.ac.uk/cbm/} and \\
\noindent\href{https://github.com/sneceesay77/mr-performance-modelling}{https://github.com/sneceesay77/mr-performance-modelling}

The main contributions of our work are:
\begin{itemize}
\item A parametric models for predicting the execution time of a big data application running on a YARN cluster.
\item A generic performance modelling approach to model the execution time of MapReduce (YARN)~\cite{vavilapalli2013apache} applications tested on different clusters and data sizes.
\item An insight into the performance characteristics of the generic phases of MapReduce. 
\item An understanding of how MapReduce applications of different design patterns perform. 
\end{itemize}

\section{Background: The MapReduce Pattern }\label{background}
\begin{figure}[h!]
 \vspace{-3mm}
 \includegraphics[width=0.15\textwidth]{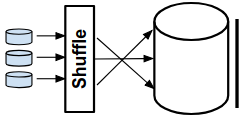}
 \centering
 \caption{\textbf{The MapReduce Communication Pattern}}
 \label{mrpattern}
\end{figure}

MapReduce \cite{dean2008mapreduce} initiated by researchers from Google \label{mrcompattern} is one of the most prevalent patterns in big data processing. The open-source community then created an open-source version of MapReduce known as Hadoop MapReduce. MapReduce 2.0 uses YARN \cite{vavilapalli2013apache} (Yet Another Resource Negotiator) for cluster resource management. In the context of YARN, a container is as a logical unit of computation assigned with CPU and memory where individual MapReduce tasks of an application are executed. They run on nodes and are managed by a Node Manager process. The node manager sends periodic reports of task progress to Application Master which monitors and manages the entire application progress. After a successful resource allocation by YARN's Resource Manager, the MapReduce execution starts by first reading data read from HDFS \cite{borthakur2008hdfs} and fed to mappers for processing. The mappers then write their intermediate results to disk. If a reducer is defined, these results are shuffled and processed by the reducers, and the final results are written to the disk. As shown in Figure \ref{mrpattern}, the time taken by a job depends on the execution of the last reducer as there is a write barrier at the end of a job. Although there are faster and newer big data processing frameworks like Apache Spark \cite{zaharia2012resilient} MapReduce is still relevant in the big data processing domain. 

\begin{figure*}
\centering
 \includegraphics[width=0.7\textwidth]{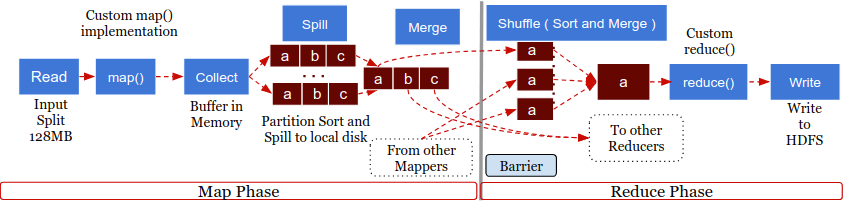}
 \caption{\textbf{A MapReduce Workflow For Each Task \cite{white2012hadoop}. The blue blocks are the different phases, whiles the brown blocks are the process data in transition from one stage to another.}}
 \label{mrworkflow}
\end{figure*}

In order to define a representative model to predict the performance of a MapReduce application, all the phases shown in in Figure \ref{mrworkflow} should be considered. The phases can be grouped into user-defined and framework-defined phases. User-defined phases are implemented by the programmer to provide customised functionality, e.g. the \textit{map()} and \textit{reduce()} functions. Framework-defined phases are defined and controlled by the framework, and users cannot modify those functionalities. Examples are \textit{read}, \textit{collect}, \textit{spill}, \textit{merge}, \textit{shuffle} and \textit{write}. 
In Section \ref{theoriticalmodels}, we will present detailed explanation for each of the phases of the MapReduce pipeline illustrated in Figure~\ref{mrworkflow}.

\section{Theoretical Models of The MapReduce Pattern}\label{theoriticalmodels}
In this chapter, we present the lessons learnt from the in-depth exploration of the MapReduce source code and its internals. 



\subsection{\textbf{Map Phase}}
As shown in Figure \ref{mrworkflow}, the Map Phase consists of the following sub-phases: \textit{Read}, \textit{Map}, \textit{Collect}, \textit{Spill} and \textit{Merge}.

\subsubsection{\textbf{Read}} This phase reads a configurable input split or data block from the \textit{Hadoop Distributed File System} (HDFS). Each record in the input split is then sent to a customised $map()$ function for processing. The cost in the read phase depends on the size of the input split.
Given a specific cluster, we can define the cost model of the read phase with a linear function of the size of the input split: 
\begin{equation}
T_{read} = f(d) = \beta_{0r} + \beta_1d
\label{eq:readphase}
\end{equation}
\noindent where $d$ represents the size of the data and $\beta_0$ and $\beta_1$ represents the unknown parameters of the linear function. However, for a given MapReduce job, the time taken to read each block of 128MB of data is usually the same.
\subsubsection{\textbf{Custom Map}}
This is the second sub-phase of the Map Phase. The map function contains a user-defined code to specify how to process records consumed from the input splits. Since this phase is user-defined, we, therefore, devised a way to approximate the time as shown below.  

 \begin{equation}
T_{map} = \frac{\sum_{i=1}^{n}T_{rec}}{M}/{N_c}
 \label{eq:mapphase}
 \end{equation}
Where $n$ is the number of records in the input split fed to the map function, $T_{rec}$ is the time taken to process each record, M is the number of mappers and $N_c$ is the number of containers. 

Moving forward, the amount of data that passes through each stage is crucial to the accuracy of the model, therefore the data emitted by the map phase $M_d$ should also be calculated as shown in Equation \eqref{eq:mapsel}. $d$ is the total input data for task $t$ and $M_{sel}$ is the map selectivity ratio
\begin{equation}
M_d = d * M_{sel} 
\label{eq:mapsel}
\end{equation}
\subsubsection{\textbf{Collect}}
The output of each map task is not directly written to disk and consumed by the reducers. Instead, they are buffered and presorted in memory. This phase also depends on the amount of data emitted by the mappers. A linear cost model of the collect phase is given as below:
\begin{equation}
T_{collect} = f(M_d) = \beta_{0c} + \beta_3M_d
\label{eq:collectphase}
\end{equation}
\subsubsection{\textbf{Spill}}
In this phase, map output data is partitioned, sorted in memory and written to local disks. Writing data to local disk is the main bottleneck. The more data to spill, the more time it will take. We can, therefore, represent the cost model using the linear relationship shown below:
\begin{equation}
T_{spill} = f(M_d) = \beta_{0s} + \beta_4M_d
\label{eq:spillphase}
\end{equation}
\subsubsection{\textbf{Merge}}
Each time the collect buffer reaches its configurable threshold, a new spill file is created.
Similar to the spill phase, the cost depends on the amount of data to merge-sort and write back to disk. From the analysis Hadoop source code, this phase uses the merge-sort algorithm which has an average complexity of $nlogn$. Since the focus of this work is to identify key performance drivers, we ignore the less significant I/O cost at this stage. Hence, the relationship to model this phase is:
\begin{equation}
T_{merge} = f(M_dlogM_d) = \beta_{0m} + \beta_5M_d(logM_d)
\label{eq:mergephase}
\end{equation}

\subsection{\textbf{Reduce Phase}}
As shown in Figure \ref{mrworkflow}, Reduce Phase of each task consists of the following sub-phases: \textit{Shuffle}, \textit{Reduce} and \textit{Write}. The cost analysis of each phase will be discussed as follows.

\subsubsection{\textbf{Shuffle}}
When all map outputs have been copied, data are then merged into larger ones while maintaining their sorting order to be consumed by the next stage. To model this phase, we calculated the data that are shuffled through to each reducer. The total data processed by each reducer can be represented as:
\begin{equation}
S_d = \frac{d * M_{sel} * M_t }{R_t} 
\label{eq:shuffledata}
\end{equation}
Where $M_t$ is the total number of mappers, and $R_t$ is the total number of reducers.
Using Equation \eqref{eq:shuffledata}, the cost model of the shuffle phase can be formulated as: 
\begin{equation}
T_{shuffle} = \beta_{0f} + \beta_6S_d + \beta_7M_t
\label{eq:shufflephase}
\end{equation}
\subsubsection{\textbf{Custom Reduce}}
For each key in a data partition, the reduce function is executed. We use the formula below to approximate the time it takes for the reduce phase to complete.

 \begin{equation}
T_{reduce} = \frac{\sum_{i=1}^{n}T_{key}}{R_t}/{N_c}
 \label{eq:reducephase}
 \end{equation}
Here, $N_c$ is the total number of containers.
\subsubsection{\textbf{Write}}
This is the last phase of the MapReduce pipeline. The output of custom reduce function is collected and written to HDFS. To model this phase, first, we modelled the total data that the reducer task emits, i.e.,
\begin{equation}
R_d = S_{d} * R_{sel} 
\label{eq:reducedata}
\end{equation}
\noindent where $S_d$ is the total shuffle data fed to a reduce task and $R_{sel}$ is the ratio of input and output sizes. Using the relations in Equation \eqref{eq:reducedata}, we can, therefore, define the cost model of the write as a linear function of the reduce output data size, which is:
\begin{equation}
T_{write} = f(R_d) = \beta_{0w} + \beta_8R_d
\label{eq:writephase}
\end{equation}
\subsection{\textbf{Combining it All Together}}
Now that we have proposed the cost model for the individual phases of the MapReduce pipeline. Putting them all together for the two main phases, we have:
\begin{equation}
T_{mt} = T_{read} + T_{map} + T_{collect} + T_{spill} + T_{merge}
\label{eq:mapt}
\end{equation}
Now substituting the phases with their corresponding linear cost models and replacing all the constants $\beta_{0*}$ as  $\beta_{0}$ together to have
\begin{equation*}
T_{mt} = \beta_0 + \beta_1d + \beta_2d + \beta_3M_d + \beta_4M_d + \beta_5M_dlogM_d + T_{map} + \epsilon
\label{eq:maptotal}
\end{equation*}
Similar to the Map Phase, we can combine all the initial formula in the sub phases of the reduce phase, i.e.,
\begin{equation}
T_{rt} = T_{shuffle} + T_{reduce} + T_{write}
\label{eq:redt}
\end{equation}
Putting the cost models together, we have: 
\begin{equation}
T_{rt} = \beta_0 + \beta_6S_d + \beta_7S_d + \beta_8R_d + T_{reduce} + \epsilon
\label{eq:reducetotal}
\end{equation}
\subsection{\textbf{Cost Model For The Entire Process}}
The models presented in the previous section represent models for a single task. MapReduce runs several tasks in parallel. Depending on the system resources, all task may run in one round, or in most cases, there will be several rounds of tasks before the entire job finishes. Before YARN, map and reduce slots were used to determine the number of tasks that can run concurrently. This approach does not fully utilise the cluster resource and previous performance models using this approach cannot be applied to YARN-based MapReduce application. YARN uses \textit{containers} for task execution. The number of containers in a cluster determines the number of tasks that can run concurrently. Assuming that a cluster has 20 containers $N_c$ and job $j$ has a total of 100 map tasks $M_t$, there will be at least five rounds of 20 map phase with each phase running 20 tasks concurrently.
Using this logic we can modify both the Map and Reduce final formulas as follows:
\begin{equation}
\begin{aligned}
T_{mt} &= \frac{M_t}{N_c}[T_{read} + T_{map} + T_{collect} + T_{spill} + T_{merge}] \\ 
&= \frac{M_t}{N_c}[\beta_0 + \beta_1d + \beta_2d + \beta_3M_d + \beta_4M_d +\\ 
&\;\;\;\;\;\;\;\;\;\;\;\;\beta_5M_dlogM_d + T_{map}] + \epsilon
\end{aligned}
\label{eq:mapjob}
\end{equation}
Similarly, the same can be done for the reduce phase:
\begin{equation}
\begin{aligned}
T_{rt} & = \frac{R_t}{N_c}[T_{shuffle} + T_{reduce} + T_{write}] \\ 
& = \frac{R_t}{N_c}[\beta_0 + \beta_6S_d + \beta_7S_d + \beta_8O_d + T_{reduce}] + \epsilon
\end{aligned}
\label{eq:reducejob}
\end{equation}
The final cost model of the entire job can be obtained by merging Equation \eqref{eq:mapjob}, Equation \eqref{eq:reducejob}, the custom phases (Equation \ref{eq:mapphase} and Equation \ref{eq:reducephase}). 
\begin{equation}
\small
\begin{aligned}
T_{job} & = [\frac{M_t}{N_c}[T_{mt}] + \frac{R_t}{N_c}[T_{rt}]] + \epsilon \\ 
& = \frac{M_t}{N_c}[\beta_0 + \beta_x\textprime d + \beta_y\textprime M_d + \beta_5M_dlogM_d + T_{map}] + \\ 
& \;\;\;\;\frac{R_t}{N_c}[\beta_0 + \beta_z\textprime S_d + \beta_8R_d + T_{reduce}] + \epsilon \\
\end{aligned}
\label{eq:totaljob}
\end{equation}

\section{Phase Profiling Methodology}\label{phaseprofilingmethodology}


The profiling process is inspired by \cite{zhang2013benchmarking}, but different in terms of the underlying system, workloads used and the experimental approach. It includes running generic benchmarks on the target cluster multiple times with various configurations. For brevity, we assume that the network is stable, no node failure and a non-shared environment and therefore we have configured YARN to use the FIFO \cite{vavilapalli2013apache} scheduling algorithm. For each phase of MapReduce, 
data is collected from the corresponding YARN logs on the cluster.


\begin{equation}
BM_{i} = (D_i, M_{sel_i}, B_{size_i})
\label{eq:genericbm}
\end{equation}
Each generic benchmark has the following parameters: $D_i$ is the size of the input dataset, $M_{sel}$ represents the map selectivity and $B_{size}$ represents the block-size. For each profiling step, we vary $D_i$ to read data size ranging from 500MB to 5GB with an interval of at most 500MB. This parameter mainly profiles the read phase. We use $M_{sel}$ to parameterise map selectivity which is the ratio of map input to the map output. It is the amount of data that proceeds to the later stages of the MapReduce pipeline. For each input data size we vary the $M_{sel}$ from 10\% to 100\% using 10\% interval.

This parameter affects the collect, spill, merge and shuffle phases. We also vary $B_{size}$ using 64MB and 128MB. For example, for a given input data size of 5GB, we executed the benchmark 20 times (10*2, where 10 is from the various value of $M_{sel}$, and 2 is from values of $B_{size}$).

\begin{figure}[h!]
\vspace{-3mm}
 \includegraphics[width=0.4\textwidth,height=1.5cm]{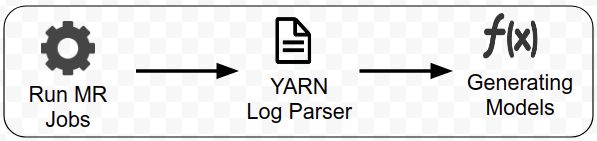}
 \centering
 \caption{\textbf{A Phase Profiling Pipeline Methodology}}
 \label{phaseprofile}
\end{figure}

As shown in Figure \ref{phaseprofile}, there are three main steps involved in the generic benchmarking setup. First, we modified the MapReduce 3.0
source code and added the necessary codes to obtain the running times of each stage. Since MapReduce uses counters to present task and job-related statistics to the user, we added six new counters to represent the six generic phases. 
The Mapper $interface$ has a logic in the $run()$ function that adheres to the map selectivity setting. The reduce function outputs the entire shuffled data. To implement the logic of map selectivity, we used the \textit{Teragen}
program provided by the MapReduce framework to generate the dataset. Each row generated by Teragen has a size of 100 bytes. Therefore, to process only 10\% of a 500MB input, we invoked Teragen to generate [(500MB to Bytes) /100] rows, and we override the $run()$ function of MapReduce interface to stop after the map had processed the 10\% threshold. As shown in Figure \ref{phaseprofile}, the second stage in the pipeline is the YARN log parser. For each job and tasks, YARN logs are collated, aggregated and parsed to extract the relevant counters and their respective values for further processing. These new data are grouped into the various phases of the MapReduce framework. The final stage of the phase profiling pipeline uses linear regression to generate the actual model parameters using the $R$ statistical programming language

\section{Generating the Actual Models and Parameters Using Regression and Cross Validation}\label{actualmodels}
\begin{figure}[h!]
 \includegraphics[width=6.5cm]{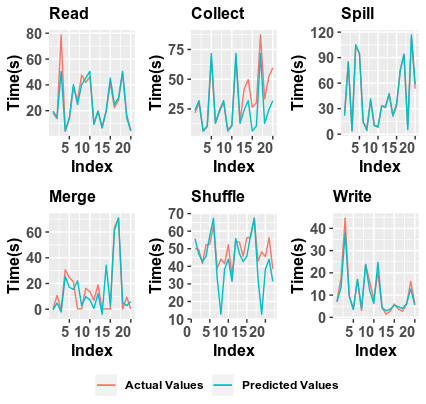}
 \centering
 \caption{\textbf{Results of 10 Fold Cross-Validation and Prediction on the 8 Nodes cluster}}
 \label{crossvalidation}
\end{figure}

Figure~\ref{crossvalidation} gives the results of the models on the processed YARN log data, using a 10 fold cross-validation on each dataset. 
The observations for each are discussed in the following subsections. The plot shows the results of the predicted and actual values. To select the right parameters and evaluate the accuracy of the model of each phase, we consider the best practices for model selection. For each model, we study the effect of $p$-value, \textit{Root Mean Squared Error} $RMSE$ \cite{barnston1992correspondence}, $Adjusted - R-squared$  \cite{miles2005r} and $Multiple-R-squared$ \cite{miles2005r}. First, to select the most important parameters of the model, we performed backwards-elimination and accept any variable whose $p$-value $\leq$ $0.05$. $RMSE$ measures the standard deviation of the residuals or prediction errors. In determining the model accuracy, we evaluate how small the value of $RMSE$ is considering the range of the dependable variable we are using. In our case, the dependent variable is \textit{execution time} represented on the Y-axis of Figure~\ref{crossvalidation}. 

\begin{figure}[h!]
 \includegraphics[width=8cm]{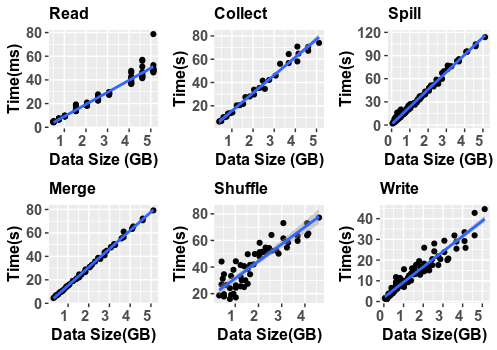}
 \centering
 \caption{\textbf{A plot and best fit for each of the generic phase: The main purpose of this plot is to show how performance(ms) changes with respect to data size (MB). 
 }}
 \label{genericmodels}
\end{figure}

Our general assumption is that in most cases the performance of these phases depends on the amount of data that it processes. This assumption is confirmed in Figure~\ref{genericmodels}. We can see that, in most cases as we increase the size of the dataset, the processing time generally increases and therefore a possible linear relationship. The last two phases in Figure~\ref{genericmodels} i.e. shuffle and write have a less linear relationship compared to the first four phases. To select the best learning method, we modelled the data using different Machine Learning (ML) approaches. As shown in Table~\ref{table:rmsersquared}, the best ML approach for both the Shuffle and Write phase is the support vector machine algorithm but not by a wide margin compared to linear regression. The algorithm has the highest R-squared value and the best RMSE in our case. Note that for equations all data measurements are in megabytes and time in milliseconds.  
\begin{table}[h!]
 \caption{RMSE and R-Squared values for Shuffle and Write}
 \captionsetup{font=scriptsize}
 \centering
  \begin{tabular}{|c|c|c|c|}
 \hline
 \textbf{Algorithm} & \textbf{RMSE(ms)} & \textbf{R-Squared} & \textbf{Phase} \\
 \hline
 SVM  & 4655.42 & 0.96 & Shuffle\\ \hline
 Random Forest  & 5027.23 & 0.95 & Shuffle \\ \hline  
 Decision Tree  & 8370.37 & 0.78 & Shuffle \\ \hline  
 Linear Regression & 4774.75 & 0.95 & Shuffle \\ \hline  
 SVM  & 2747.380 & 0.94 & Write\\ \hline
 Random Forest   & 2940.09 & 0.93 & Write \\ \hline  
 DecisionTree   & 9366.400  & 0.68 & Write \\ \hline  
 Linear Regression  & 6256.47 & 0.88 & Write \\ \hline  
\end{tabular}
 \label{table:rmsersquared}
 \vspace{-0.5cm}
\end{table}

\subsection{\textbf{Read Model}}
The Read Phase of Figure \ref{genericmodels} supports our initial assumption in \eqref{eq:readphase} that there is a linear relationship between the size of input data and the processing time. The model obtained using linear regression is presented in Equation \eqref{eq:readmodel}. The read plot in Figure \ref{crossvalidation} and $RMSE$ value of $4.08$ proves a strong case for the accuracy of the model. 
\begin{equation}
T_{read} = 0.01 \times D + 1.33
\label{eq:readmodel}
\end{equation}
\subsection{\textbf{Collect Model}}
Map output data is collected as soon as map tasks complete; once the circular buffer of each map task reaches the predefined threshold of 80\% data is then written to disk. This behaviour can be observed from Figure \ref{genericmodels}, a spike in the amount of time taken is at 80\% $M_{sel}$. The model for the collect phase in our test environment is illustrated as below:
\begin{equation}
T_{collect} = 0.01 \times M_{sel} + 0.97
\label{eq:genericcm}
\end{equation}
Again, the $RMSE$ value of $3.036$ and the Collect plot in Figure \ref{crossvalidation} supports our model.
\subsection{\textbf{Spill Model}}
As shown in Figure \ref{genericmodels}, our linearity assumption between data size and processing time is confirmed. The model obtained after the cross-validation is presented in Equation \eqref{eq:genericsm}. The model has $RMSE$ value is $2.522$ which would be considered a good model in this case.
\begin{equation}
T_{spill} = 0.02 \times M_{sel} + 0.98
\label{eq:genericsm}
\end{equation}
\subsection{\textbf{Merge Model}}
Spill files are merged together to form bigger files. From the plot, we can see a close relation between the actual and the predicted values with an $RMSE$ of $9.36$. The parameters of the model are shown in Equation \ref{eq:genericmmodel}
\begin{equation}
T_{merge} = 0.002 \times M_{sel}logM_{sel} + 4.80
\label{eq:genericmmodel}
\end{equation}
\subsection{\textbf{Shuffle Model}}
This phase is the most expensive of all because it involves inter-node communication. The model main parameters are the amount of data being shuffled and the number of Mappers task. 
The $RMSE$ value obtained after is $4655$. 
\begin{equation}
T_{shuffle} = 10.45 \times S_d + 579.48 \times M_t + 6144.6\footnote{Linear equation, however we used the SVM model in the prediction.} 
\label{eq:genericshfmodel}
\end{equation}
\subsection{\textbf{Write Model}}\label{writemodel}
From Figure \ref{genericmodels} (Write Phase), the fitted line shows a less-linear relationship between the amount of data and the time taken to write that data to HDFS. The cross-validation plot also shows a strong relationship when the test values are used in the predictive model. The model has an acceptable $RMSE$ of $2427$
\begin{equation}
T_{write} = 6.94 \times R_d + 2139.98
\label{eq:genericwritemodel}
\end{equation}
\subsection{\textbf{Custom Map and Reduce}}
Since these phases have a custom implementation that depends on the problem being solved, we, therefore, used Equation \eqref{eq:mapphase} and Equation \eqref{eq:reducephase} for approximation.
\subsection{\textbf{Generating the Input Parameters}}
With the parameters from profiling, $T_{job}$ in \eqref{eq:totaljob} is:
\begin{equation*}
\small
\begin{aligned}
&\frac{M_t}{N_c}[(0.01 \times D + 1.33)+T_{map}+(0.01 \times M_{sel} + 0.97)+\\ 
&\;\;\;\;\;\;\;(0.02 \times M_{sel} + 0.98)+(0.002 \times M_{sel}logM_{sel} + 4.80)] + \\ 
& \;\;\;\;\;\;\frac{R_t}{N_c}[(10.45 \times S_d + 579.48 \times M_t + 6144.6)+T_{reduce}+\\
& \;\;\;\;\;\;\;\;\;(6.94 \times R_d + 2139.98)] + \epsilon 
\end{aligned}
\label{eq:totaljobcomplete}
\end{equation*}
Users now need to provide the actual values for the input parameters listed in Table~\ref{table:logextracted}.
\begin{table}[h!]
 \caption{Metrics Extracted From Logs}
 \centering
  \begin{tabular}{|c|c|c|}
  \hline
 \textbf{Name} & \textbf{Variable} & \textbf{Description} \\
\hline
 Num Of Bytes Read & $D$ & Total Size of Data \\  \hline
 Map Output Bytes & $M_d$ & Data Outputted by Mappers \\  \hline
 Map Selectivity & $M_{sel}$ & $(M_d/d) * 100$ \\  \hline
 Bytes Shuffled & $S_d$ & Data shuffled, refer to Eq: \eqref{eq:shuffledata} \\  \hline
 Bytes Written & $R_d$ & Data Outputted by Reducers \\  \hline
 Total Mappers & $M_t$ & $\frac{D}{BlockSize}$ \\  \hline
 Total Reducers & $R_t$ & Optimisable \\  \hline
 Number of Containers & $N_c$ & Inferred from Cluster Config. \\  \hline
 Map Time & $T_{map}$ & {Total time map() fns} \\  \hline
  Reduce Time & $T_{map}$ & {Total time reduce() fns} \\  \hline
\end{tabular}
 \label{table:logextracted}
\end{table}
To get the values for each of these parameters, the first step would be to run an application on the profiled cluster using a representative input dataset for the application. The application logs are then parsed to the YARN log parser to collect the values of the parameters for substitution into the corresponding models. This would provide the input parameters to generate the final predictive model. 
\begin{table}[h!]
\caption{The values for each parameters for Reduce Side Inner Join algorithm extracted from YARN logs.}
 \centering
  \begin{tabular}{|c|c|}
  \hline
 \textbf{Variable} & \textbf{Value} \\
 \hline
 $D$ & 19584 \\ \hline  
 $M_t, R_t$ & 153,11 \\ \hline  
 $T_{map},T_{reduce} (ms)$ & 33069,286257 \\ \hline  
 $M_d$ & 128 \\ \hline  
 $M_{sel}$ & $100\%$ \\ \hline  
 $S_d$ & 19584  \\ \hline  
 $R_d$ & 19584 \\ \hline  
 $N_c$ & 8 \\ \hline  
\end{tabular}
 \label{table:runningex}
 \vspace{-3mm}
\end{table}
To illustrate, we take the Reduce Side Inner Join algorithm as a target application. The generated values for each of the input parameters are listed in Table~\ref{table:runningex}. To facilitate the generation of the values from YARN logs we provided Java and R scripts available on our Github page.
\section{Experiment and Evaluation}\label{experimentandevaluation}
\subsection{\textbf{Setup}}
In order to gauge the applicability of our approach on an arbitrary setup, we conducted two sets of experiments on two different hardware setups. On the first setup, we used a single node YARN cluster with 32GB memory and 8 CPUs. YARN was allocated 24GB and a minimum of 3GB per container
In the second setup, we used an 8-node in-house YARN cluster to mimic a real-world deployment scenario. Each of these nodes has 8GB of RAM, 8 vCPUs and a 500GB of storage space. YARN is allocated 48GB of memory, and a maximum of 8 containers can run at a time. From the results of the two setups for the experiments, we have a strong case to conclude that, the approach can be used to profile an arbitrary cluster size. 
In all the experiments, we have adapted the algorithms discussed in the design patterns book \cite{Miner:2012:MDP:2632825} and used the same $StackOverflow$ data source but a more updated version. Table \ref{table:algorithms} summarises all the workloads and their data-sizes used in this work. To have a representative cost model for the various MapReduce phases and the entire process, we studied the most common design pattern algorithms of MapReduce presented in \cite{Miner:2012:MDP:2632825}. The motivation of using MapReduce design pattern approach in the experimental setup is to avoid bias towards specific applications only. 
As shown in Table \ref{table:algorithms}, we have only included large datasets in the experiment to narrow our focus on big data and data-intensive applications. We have identified four common design patterns, and for each of these patterns, we tested at least three algorithms to show how algorithms using the same pattern relates to one another. We assume that algorithms with the same pattern would have similar performance characteristics.
\begin{itemize}
\item Summarisation Pattern: This pattern provides a summary of an input dataset. The popular MapReduce program is a good example. 
\item Filtering Pattern: This pattern filters and returns a subset of a given original dataset. In most cases, there is a reduction in the amount of output dataset compared to the input dataset. Example of algorithms in this pattern are Distributed $Grep$, $Distinct$, and $Top$ $K$.
\item Data Organisation Pattern: This pattern deals with the reorganisation from one structure to another. Example, transforming table data to JSON structure.
\item Join Pattern: This pattern processes related data stored in the input files. The most popular type of Join in MapReduce is the Reduce side join. It works in all cases but can be slow as the size of the data increases.
\end{itemize}
\begin{table}[h!]
 \caption{Workloads Used}
 \centering
  \begin{tabular}{|c|c|c|}
\hline
 \textbf{Algorithm} & \textbf{Design Pattern} & \textbf{Data Size} \\
 \hline
 MinMaxCount & Summarisation & 16GB \\ \hline  
 Average Count & Summarisation & 16GB \\ \hline  
 Median and Std. & Summarisation & 16GB \\ \hline  
 Inverted Index & Summarisation & 12GB \\ \hline  
 Grep & Filtering & 12GB \\ \hline  
  Top X & Filtering & 2.7GB \\ \hline  
 Distinct & Filtering & 16GB \\ \hline  
 Structure to Hierarchy & Data Organisation & 12GB,16GB \\ \hline  
 Total Order Sorting & Data Organisation & 2.6GB \\ \hline  
 Shuffling & Data Organisation & 2.7GB \\ \hline  
  RSJ Inner & Join & 16,2.7GB \\ \hline  
 RSJ Left Outer & Join & 16,2.7GB \\ \hline  
 RSJ Right Outer & Join & 16,2.7GB \\ \hline  
 RSJ Full Outer & Join & 16,2.7GB \\ \hline  
\end{tabular}
 \label{table:algorithms}
 \vspace{-3mm}
\end{table}
\subsection{\textbf{Evaluation of Results}}
For each of the two setups, the same experiments are repeated and the results are plotted and discussed side by side. As expected, the time taken by the single node cluster for each of the algorithms is mostly greater than the eight node cluster. Below, we discuss the rest of the results in details. 
\subsubsection{Summarisation Pattern}
\begin{figure}[h!]
 \includegraphics[width=0.4\textwidth, height=4cm]{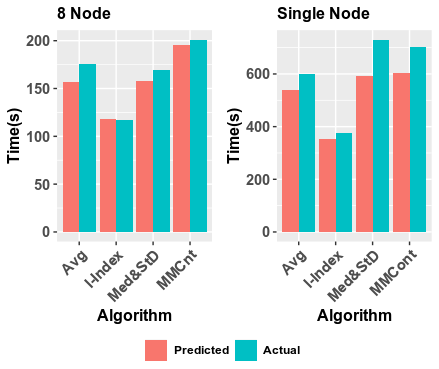}
 \centering
 \caption{Summarisation Pattern}
 \label{plotsummarisation}
\end{figure}

The algorithms and the results obtained included are listed in \ref{table:summarisation} and corresponding plot in Figure~\ref{plotsummarisation}. The percentage prediction error for both setups is less than 16\%. Also note that the algorithms in this pattern have a similar completion time boundaries.
\begin{table}[h!]
 \caption{Results for Summarisation on 8-Node Cluster}
\captionsetup{font=scriptsize}
 \centering
  \begin{tabular}{|c|c|c|c|}
\hline
 \textbf{Algorithm} & \textbf{Predicted (sec)} & \textbf{Actual (sec)} & \textbf{Error\%} \\
 \hline
 MinMaxCount & 196 & 201 & 2\\  \hline
 Inverted Index & 118 & 117 & -1 \\  \hline
 Average Count & 157 & 176 & 11\\  \hline
 Median and Std. Dev & 158 & 169 & 7\\  \hline
\end{tabular}
 \label{table:summarisation}
 \vspace{-4mm}
\end{table}
\subsubsection{Filtering Pattern}
In filtering, we included, Grep, Distinct and Top 100 algorithm. 
Table \ref{table:filtering} shows the results of the experiment and the respective errors of each algorithm. We have also observed that our prediction error is less than 14\% from the observed values for both setups. 
\begin{figure}[h!]
 \includegraphics[width=0.4\textwidth, height=4cm]{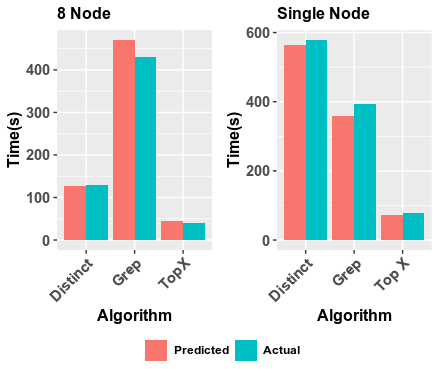}
 \centering
 \caption{Filtering Pattern}
 \label{plotfiltering}
\end{figure}

\begin{table}[h!]
 \caption{Results for Filtering Pattern on 8-Node Cluster}
 \setlength\tabcolsep{1.5pt}
 \captionsetup{font=scriptsize}
 \centering
  \begin{tabular}{|c|c|c|c|}
 \hline
 \textbf{Algorithm} & \textbf{Predicted (sec)} & \textbf{Actual (sec)} & \textbf{Error\%} \\ \hline
  Grep & 470 & 430 & -9\\   \hline
 Top X & 45 & 40 & -12 \\ \hline  
 Distinct & 128 & 130 & 2 \\ \hline  
\end{tabular}
 \label{table:filtering}
\end{table}
\subsubsection{Data Organisation Pattern}
\begin{figure}[h!]
 \includegraphics[width=0.4\textwidth, height=4cm]{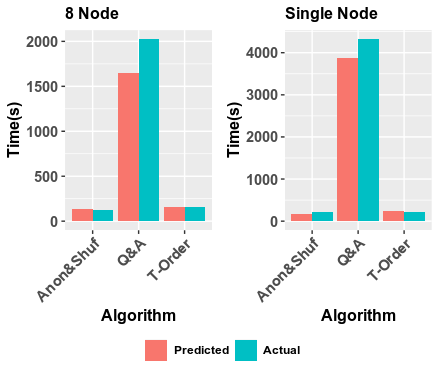}
 \centering
 \caption{Data Organisation Pattern}
 \label{plotdataorganisation}
\end{figure}

In data organisation, we have observed that the amount of input data is mostly the same as the amount of output data. Here data is just reorganised and there is no data pruning component. The Question and Answer Hierarchy algorithm merged data from two big data sets. For each Question posted in Stackoverflow, the corresponding answers are collated from the Post file. The results of the three executed experiments are illustrated in Figure \ref{plotdataorganisation} and Table \ref{table:dataorganisation}.
\begin{table}[h!]
 \caption{Results for Data Organisation Pattern}
 \captionsetup{font=scriptsize}
 \centering
  \begin{tabular}{|c|c|c|c|}
\hline
 \textbf{Algorithm} & \textbf{Predicted (sec)} & \textbf{Actual (sec)} & \textbf{Error\%} \\
 \hline
 Q\&A Hierarchy & 1653 & 2024 & 18 \\  \hline
 Total Order & 162 & 160 & -1\\  \hline
 Anonymise \& Shuffling & 131 & 125 & -8 \\  \hline
\end{tabular}
 \label{table:dataorganisation}
\end{table}
\subsubsection{Join Pattern}
\begin{figure}[h!]
 \includegraphics[width=0.4\textwidth, height=4cm]{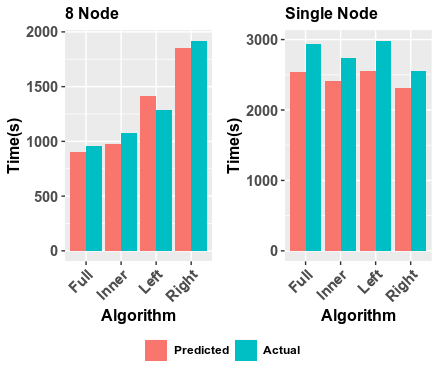}
 \centering
 \caption{Join Pattern}
 \label{join}
\end{figure}

Joins are one of the most expensive operations in MapReduce, this is evident in our experimental results. We executed inner, left-outer, right-outer and full-outer joins using the Reduce Side Join Algorithm. We used the Post and Comments dataset curled from Stack Overflow. 
The prediction error for both setups is at most 10\% of the observed value. 
The results are shown in Table \ref{table:join} and Figure \ref{join}.

\begin{table}[h!]
\caption{Experiment Data for Join Pattern}
\captionsetup{font=scriptsize}
 \centering
  \begin{tabular}{|c|c|c|c|}
 \hline
 \textbf{Algorithm} & \textbf{Predicted (sec)} & \textbf{Actual (sec)} & \textbf{Error\%} \\
 \hline
 Reduce Side Inner & 975 & 1080 & 10 \\  \hline
 Reduce Side L-Outer& 1410 & 1285 & -10 \\  \hline
 Reduce Side Right Outer& 1854 & 1920 & 3 \\  \hline
 Reduce Side Full Outer& 900 & 960 & 6 \\  \hline
\end{tabular}
\label{table:join}
\end{table}

\subsection{Discussion} Here we revisit our three research questions (RQs) and evaluate how far we have answered them. In answering RQ0, we adopted a generic benchmarking approach by running dummy MapReduce workload on the cluster. The main idea behind this is to avoid benchmarking bias when only a particular set of workloads are used. In addressing RQ1, we have realised that out of the many variables, input parameters such as the number of mappers, number of reducers, number of containers and the amount of data that passes through each stage are the key performance drivers. The final model building process involves feeding these key performance drivers to ML algorithms. In addressing RQ2, we scaled our experimental environment from a single node setup to an eight-node cluster. We have observed that the approach we have used can be replicated to arbitrary cluster size.  

\section{Related Work}\label{relatedwork}
Standard benchmarks such as TPC-DS \cite{nambiar2006making} has been used in the research community to evaluate the performance of decision support systems. 
Big data Benchmarking and Performance modelling recently got lots of attention from researchers \cite{huang2010hibench, dongarra1979linpack,herodotou2011hadoop,glushkova2017mapreduce,verma2011resource}. \citet{huang2010hibench} and \citet{wang2014bigdatabench} developed benchmark suites for Hadoop, Spark and Streaming Frameworks. These suites consist of a set of workloads organised into related groups. These workload groups span from basic statistics, machine learning, graph processing and SQL. However, this approach makes it inefficient and cumbersome to test new workloads. Also, as stated in \citet{ceesay2017papb}, deploying these tools can be cumbersome for non-technical individuals. Therefore it becomes a challenge for wider adoption. \citet{zhang2013benchmarking}, presents a MapReduce performance model that measures generic phases of the framework. However, their work focused on an older version of Hadoop (0.20.0) which uses mapper and reducer slots for job processing. The current MapReduce framework uses YARN which efficiently manages cluster resources. Given this fact, it is clear that the approach and the model they used cannot be representative of the current MapReduce paradigm. Secondly, our experimental approach differs as well, while they pick any big data application, we picked ours grouped by the algorithms design patterns, the results of which show some interesting correlation in terms of their performance. Applications in the same pattern mostly have similar performance characteristics in terms of their execution time.
\citet{verma2011aria} proposed ARIA, a job and resource scheduler for the MapReduce framework that aims to allocate the right amount of resource to meet required service level objectives (SLOs). In their work, they extracted information from MapReduce logs as a basis for their framework. Like \cite{zhang2013benchmarking}, they used Hadoop 0.20.0, and therefore their approach will not be applicable for Hadoop 2.0 or later versions. \citet{venkataraman2016ernest}, built performance models based on a small sample of data and predicting on larger datasets and cluster sizes. However, they focus on a small subset of machine learning algorithms. \citet{popescu2012same} introduced an approach for predicting the runtime JAQL queries which focuses on mainly JSON and related data.     

\section{Conclusion and Future Work}\label{conclusion}
In conclusion, we have shown that we can model the performance of big data application in the MapReduce pattern. 
We have shown that algorithms in the same pattern tend to have similar performance characteristics when the data size is similar. 
We have also proposed a generic benchmarking approach that can be used to get performance characteristics of the various stages in a MapReduce pipeline. Using the data generated from the benchmarking, linear regression and cross-validation were used to build and validate the models for each phase. These models were used to predict the competition time for a new application, and the results were promising. In future work, we will work on modelling other communication patterns proposed in \cite{chowdhury2012coflow} such as Data Flow with Cycles, which performs computations and subsequent transformations in memory until the user actively persists them to disk. We will also consider validating our approach with a much larger cluster and datasets.
Finally, since scheduling and scalability are an integral part of big data systems, we will also study their effects on the performance of our modelling approach. 
\section{Acknowledgement}
This research is funded by EPSRC EP/R010528/1 and IsDB.
\balance

\ifCLASSOPTIONcompsoc
\else
\fi
\label{Bibliography}

\bibliographystyle{plainnat} 
\footnotesize
\bibliography{Bibliography} 

\end{document}